\documentclass[modern]{aastex62}
\received{}
\revised{}
\accepted{}
\submitjournal{}

\shorttitle{Formation of Clumpy CO and \ion{C}{1} clouds}
\shortauthors{Tachihara et al.}

\begin{document}

\title{Synthetic observations of highly clumped formation of CO and \ion{C}{1} clouds in the turbulent interstellar medium; Implications on the seed CO clouds and the CO-dark H$_{2}$ gas}

\correspondingauthor{Kengo Tachihara}
\email{k.tachihara@a.phys.nagoya-u.ac.jp}

\author[0000-0002-1411-5410]{Kengo Tachihara}
\affiliation{Department of Physics, Nagoya University, Chikusa-ku Nagoya 464-8602, Japan}

\author{Yasuo Fukui}
\affiliation{Department of Physics, Nagoya University, Chikusa-ku Nagoya 464-8602, Japan}

\author[0000-0003-0324-1689]{Takahiro Hayakawa}
\affiliation{Department of Physics, Nagoya University, Chikusa-ku Nagoya 464-8602, Japan}

\author{Tsuyoshi Inoue}
\affiliation{Department of Physics, Nagoya University, Chikusa-ku Nagoya 464-8602, Japan}



\begin{abstract}
We carried out synthetic observations of a theoretical model of interstellar carbon in CO, \ion{C}{1} and \ion{C}{2} including H$_{2}$ formation.
The theoretical model based on numerical simulations of \ion{H}{1} gas converging at a velocity of 20\,km\,s$^{-1}$ by \citet{2012ApJ...759...35I} is employed in the present study.
The gas model, which is fairy realistic and consistent with the observations of the local gas located at high $b$, is turbulent and inhomogeneous with magnetic field.
We focus on the evolution of the rare species, CO, \ion{C}{1} and \ion{C}{2} from 0.3\,Myr to 9\,Myr since the converging flows collided.
In the early phase within 1\,Myr, the Cold Neutral Medium (CNM) clumps of sub-pc scales are the formation sites of H$_{2}$, CO and \ion{C}{1}, and in the late phase from 3\,Myr to 9\,Myr the CNM clumps merge together to form pc-scale molecular clouds.
The results show that \ion{C}{1} and CO are well mixed at pc scales, and mark a significant disagreement with the classic picture which predicts layered \ion{C}{1} and CO distributions according to $A_{V}$.
We suggest that the faint CO clouds in the Oph North \citep{2012ApJ...754...95T} and the Pegasus loop \citep{2006ApJ...642..307Y} are promising candidates for such small CO clouds.
The total mass budget of the local ISM indicates the fraction of \ion{H}{1} is about 70\% and the CO-dark H$_{2}$ is small, less than a few percent, contrary to the usual assumption.
\end{abstract}



\section{Introduction}
The interstellar medium (ISM) plays an important role in galactic evolution by forming stars.
A detailed study of the ISM is possible only for the local ISM which is distributed within 300\,pc of the sun, where no giant molecular clouds (GMCs) forming high mass stars are found.
The present work focuses on the local ISM whose peak H column density is $\sim 10^{21}$\,cm$^{-1}$.
The local gas is relatively diffuse and contains little CO clouds except for dark clouds including Taurus, Chameleon, R CrA, etc.

The interstellar molecular clouds are the almost exclusive site of star formation.
As an evolutionary step prior to star formation, molecular cloud formation is an important process in galaxy evolution, and considerable efforts have been devoted to better understanding molecular cloud formation in the previous studies \citep[for a recent review see][and references therein]{2014prpl.conf....3D}.
Our understanding on cloud formation is yet preliminary both theoretically and observationally and more efforts are to be made toward confronting theories and observations and to acquire comprehensive understanding of the physical and chemical processes and their observable signatures.

Some theoretical works were made aiming at understanding formation of the molecular clouds at various scales from sub-pc to kpc \citep{2014prpl.conf....3D}.
The aim of the present work is to better understand the formation process at 0.02\,pc to 10\,pc scales by incorporating the micro physical processes including photoelectric heating and atomic and molecular cooling, cosmic ray ionization, molecular formation of H$_{2}$ and CO, and the interstellar radiation field and cosmic rays.
For the purpose, we employ one of the most advanced numerical simulations of the realistic ISM having highly turbulent inhomogeneous distribution with magnetic field.

Since we are interested in the behavior of carbon bearing species, the role of the radiation field is crucial.
The effects of the UV radiation appear as regions with diverging ionization states which are governed by $A_{V}$ under a given radiation field.
A pioneering work by \citet{1985ApJ...291..722T} developed the photo dissociation region (PDR) model, where the ISM is exposed to external UV radiation field of various intensity.
Carbon behavior, the major constituent of the ISM, attracted a broad interest and a number of theoretical simulations were undertaken previously \citep[e.g.][]{2010MNRAS.404....2G}.
Most of these studies were made for uniform or non-uniform but smooth, density distribution.
The results predicted that the \ion{C}{2}, \ion{C}{1} and CO are distributed in layers in a sequence of $A_{V}$ \citep{1985ApJ...291..722T}.
For small $A_{V}$ less than 0.3\,mag, \ion{C}{2} is dominant and \ion{C}{1} increases with $A_{V}$.

The observations of \ion{C}{1} are not so plentiful.
Large scale \ion{C}{1} observations were made with the Mt.\ Fuji sub-mm telescope and provided \ion{C}{1} distribution in nearby clouds including Taurus, Orion, etc.\ \citep{1999ApJ...524L.129M,1999ApJ...527L..59I,2001ApJ...558..176O,2003ApJ...589..378K}.
Their results showed that \ion{C}{1} and CO distributions are fairly similar and the distinct difference as predicted by the PDR model was not observed.
\citet{2015MNRAS.448.1607G} proposed highly turbulent field in the ISM in order to solve the discrepancy and showed that the turbulence can introduce more UV photons deeply into a cloud, and thereby realizes ionization over a cloud, producing similar \ion{C}{1} and CO distributions. 

The aim of the present work is to clarify the behavior of the ISM in particular the observed properties of the CO and \ion{C}{1} and their relationship with the density distribution of the ISM.
For the purpose, we employ the converging \ion{H}{1} flow model by \citet{2012ApJ...759...35I}.
This model is distinguished from others which assume high density as the initial condition; e.g., \citet{2007ApJS..169..239G} assumes uniform and static \ion{H}{1} with $\mbox{density}=100$\,cm$^{-2}$ as the initial condition, which excludes the Warm Neutral Medium (WNM).
\citet{2018ApJ...860...33F} used the model by \citet{2012ApJ...759...35I} and synthesized \ion{H}{1} and H$_{2}$ distributions.
These authors showed that the Cold Neutral Medium (CNM) has a highly clumpy distribution with a volume filling factor of 3.5\% which is consistent with the observations.
The present paper makes synthetic observations of CO and \ion{C}{1} which are included in the same model, and aims to gain better insight into the interstellar carbon behavior.
Section \ref{sec:models} gives the model description, and Section \ref{sec:results} presents details of the results.
Discussion is given in Section \ref{sec:discussion}, and Section \ref{sec:conclusions} summarizes the conclusions.

\section{Models  and the synthetic observations}\label{sec:models}
The present simulations of the converging \ion{H}{1} flows at 20\,km\,s$^{-1}$ were made by \citet{2012ApJ...759...35I}, and adopt initial conditions which consider the realistic inhomogeneous density distribution as a natural consequence of the thermal instability based on magneto-hydrodynamical code. Heating by the background FUV field and cosmic rays are taken into account as well as molecular and atomic coolings. Chemical reactions including those relevant to H$^+$, H$_{2}$, CO, \ion{C}{1} and \ion{C}{2} are considered.
More details are found in \citet{2012ApJ...759...35I}.
All the datasets have pixels of $512^{3}$ for a length of 20\,pc in each axis.
The following four models are used in the present study, while usually we showed the results of Model A in the present paper.

Model A is based on the numerical simulations which were employed in \citet{2018ApJ...860...33F}, where the \ion{H}{1} flows continuously enter into the simulation box at 20\,km\,s$^{-1}$ from the both sides along the $x$-axis.
Magnetic field is assumed to be parallel to the $x$-axis, so that the converging flows compressed very efficiently to form molecular cloud taking 10 Myr.
See more details in \citet{2012ApJ...759...35I}.

Unlike Model A, Models B, C, and D make the \ion{H}{1} flows injected from both sides along the $x$-axis collide only once around $x = 10$ pc. Model B assumes the initial velocity as follows;
\begin{eqnarray*}
V_x\,\mbox{(km\,s$^{-1}$)} & = & \frac{5x}{\mbox{(pc)}} \; \mbox{(for $x<10$\,pc)}, \\
V_x\,\mbox{(km\,s$^{-1}$)} & = & \frac{5x}{\mbox{(pc)}}-100 \; \mbox{(for $x>10$\,pc)}.
\end{eqnarray*}
This setup allows us to simulate that the ISM in the numerical domain is compressed by the shock and after that the increased pressure of the ISM is taken back to an average value of the ISM due to the cooling (in contrast, the shock compressed region in Model A keeps high thermal pressure because the converging flows are continuously set at the boundary). 
With this initial velocity field, the total kinetic energy of $\sim 2\times10^{49}$ erg is injected into the numerical domain of 20 pc$^3$ volume, which is comparable to that given by a supernova explosion. 
The field direction is taken at 45 degrees relative to the $x$-axis in the $x$-$y$ plane, while Model A assumed this angle to be 0 degrees.

Model C has smaller initial velocity as follows;
\begin{eqnarray*}
V_x\,\mbox{(km\,s$^{-1}$)} & = & \frac{2x}{\mbox{(pc)}} \; \mbox{(for $x<10$\,pc)}, \\
V_x\,\mbox{(km\,s$^{-1}$)} & = & \frac{2x}{\mbox{(pc)}}-40 \; \mbox{(for $x>10$\,pc)}.
\end{eqnarray*}

Model D has the initial velocity same as Model C, but the field direction is taken in the $x$-axis.

Synthetic CO, \ion{C}{1} and \ion{C}{2} profiles were calculated by using a line radiation transfer code RADMC-3D \citep{2012ascl.soft02015D}.
The code first computes the level populations consistent with the local density, temperature and velocity field by adopting the Large Velocity Gradient (LVG) method (also known as ``Sobolev approximation''), and then performs a ray-tracing line transfer calculation with the derived level population.
The molecular/atomic data required for the calculations (energy levels, the Einstein $A$ coefficients and rate coefficients of collisional excitation/de-excitation) are taken from Leiden Atomic and Molecular Database (LAMDA) \citep{2005A&A...432..369S}.

\section{Results}\label{sec:results}

\subsection{Density and temperature distribution}
\begin{figure*}
\includegraphics{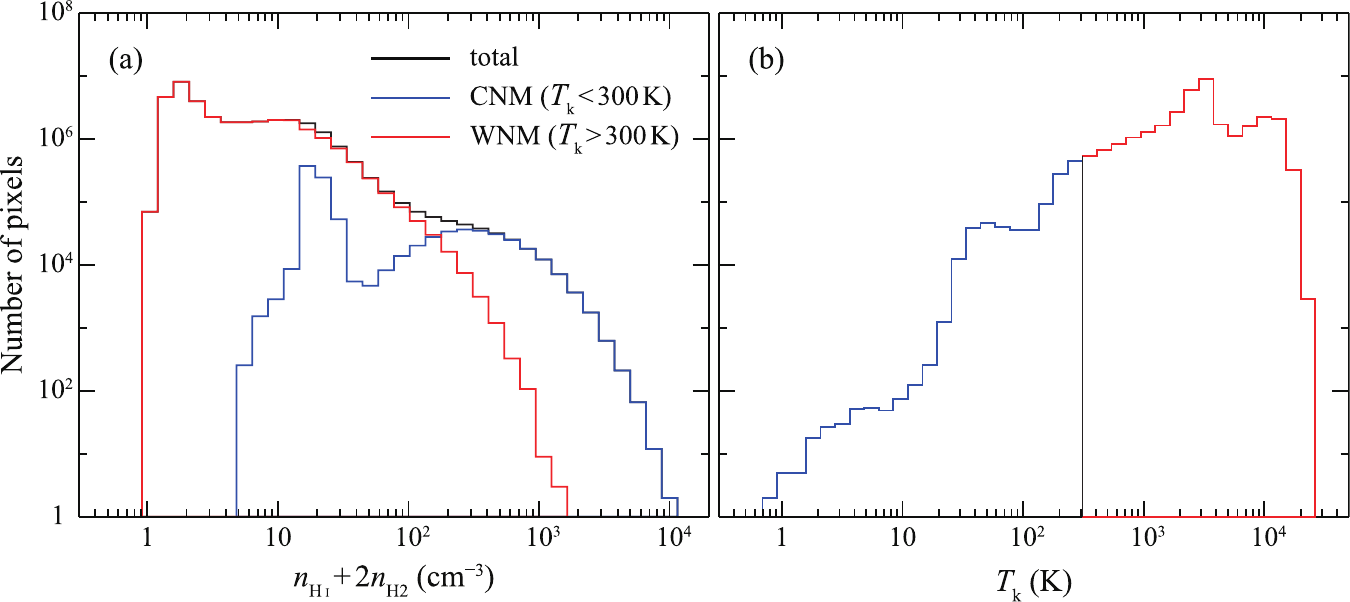}
\caption{
Histograms of (a) total hydrogen density ($n+2n_\mathrm{H2}$) and (b) kinetic temperature ($T_\mathrm{k}$) for each pixel in Model A at 0.5\,Myr \citep[reproduced from Figure 5 of][]{2018ApJ...860...33F}.
The blue lines represent the contribution of $T_\mathrm{k}<300$\,K and red lines represent that of $T_\mathrm{k}<300$\,K.
\label{fig:histo_n+T}}
\end{figure*}

\begin{figure}
\includegraphics{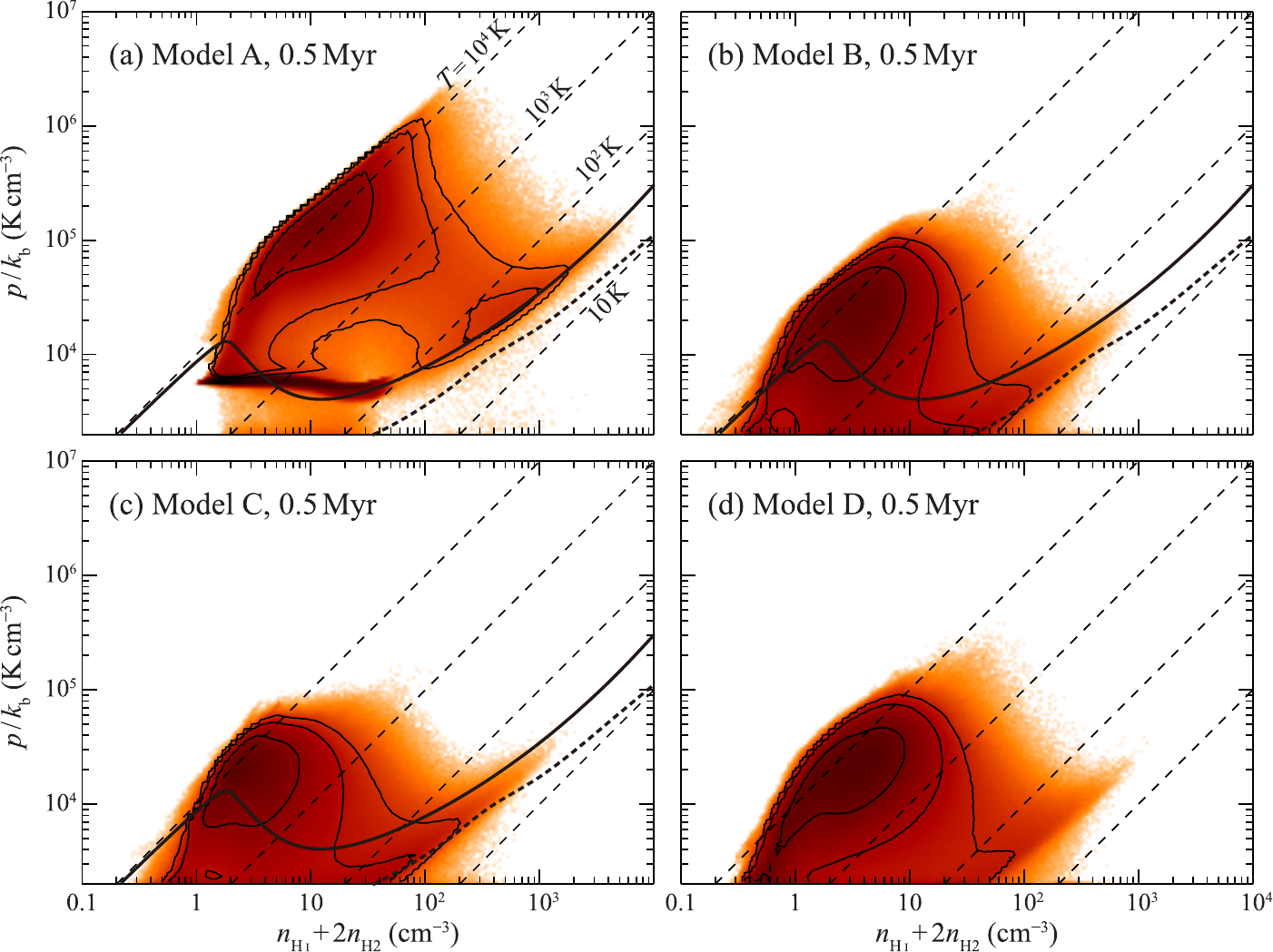}
\caption{
(a) Probability distribution function in the ($n_\mathrm{HI}+2n_\mathrm{H2}$)-$p/k_\mathrm{b}$ plane for Model A at 0.5\,Myr.
The condensation at $p/k_\mathrm{b}\sim 7\times 10^{3}$\,K\,cm$^{-3}$ shows the initial \ion{H}{1} flows prior to the collision.
The thick solid and dashed curves show thermal and chemical equilibrium curve for media of $A_{V}=0$\,mag and $A_{V}=1$\,mag \citep[reproduced from Figure 1 of][]{2012ApJ...759...35I}.
The thin dashed lines are isotherms of $T_\mathrm{k}=10$, $10^{2}$, $10^{3}$ and $10^{4}$\,K.
(b)--(d) Same as (a) but for Models B--D at 0.5\,Myr.
\label{fig:npdiagram}}
\end{figure}

Figure \ref{fig:histo_n+T} shows distributions of density and temperature for Model A at 0.5\,Myr \citep[reproduced from][]{2018ApJ...860...33F}.
The gas temperature ranges from 10\,K to $10^{4}$\,K, where the WNM and CNM have a boundary at 300\,K, and density ranges from 1 to $10^{4}$\,cm$^{-3}$.
The mass ratio of the two phases are $\sim 4:6$ for the CNM and WNM.
The pressure distributions for the present 4 models at 0.5\,Myr
are shown as a function of density in Figure \ref{fig:npdiagram}, where two stratified gas components corresponding to the WNM and CNM are seen. 
The pressure range is also quite broad from $\sim 10^{3}$\,K\,cm$^{-3}$ to $\sim 10^{6}$\,K\,cm$^{-3}$, because the medium is far from the dynamical equilibrium at a uniform pressure.
This is characteristics to the highly turbulent state, and causes a number of transient structures of \ion{H}{1} in the order of $10^{4}$\,yr as shown later. 
Small differences between the 4 models can be seen in Figure \ref{fig:npdiagram}. Model A makes higher pressure of the WNM caused by the continuous gas flow than the other 3 models. On the other hand, the pressure and density of CNM do not vary a lot among the different models. Thus the different model setups causes little effects on the results of the chemical characteristics, mass ratio, and the volume filling factors of the CNM.

\subsection{Relative abundance of carbon-bearing species, CO, \ion{C}{1} and \ion{C}{2}}
\begin{figure}
\includegraphics{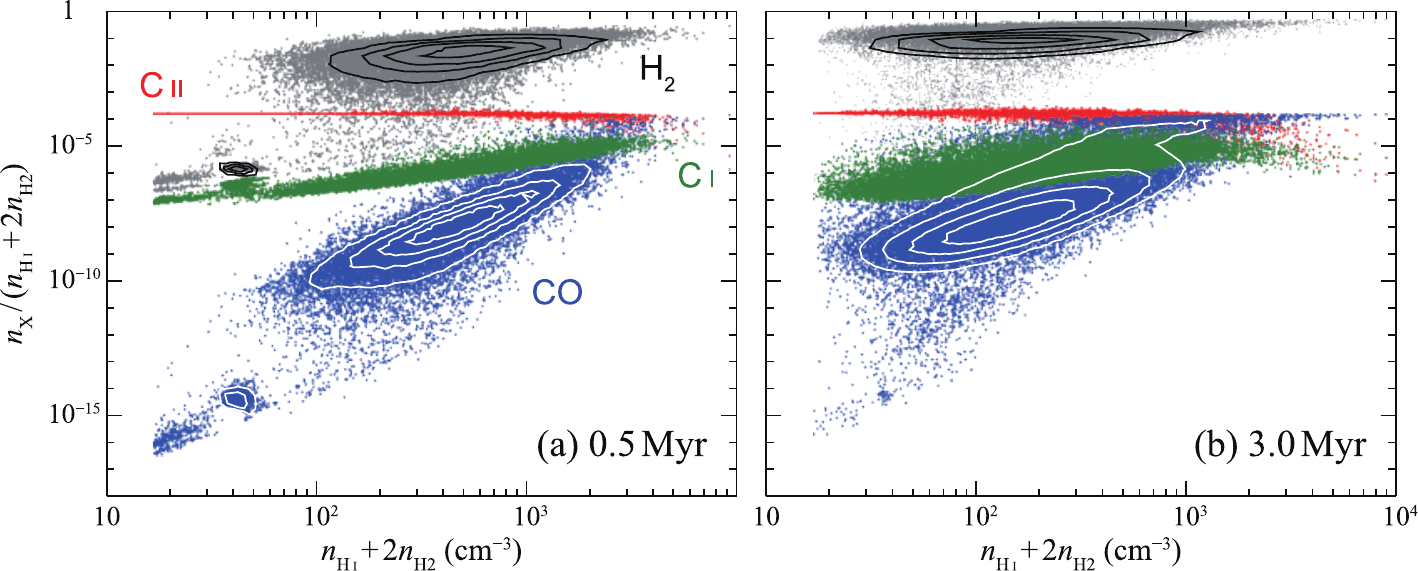}
\caption{
Relative abundance of H$_{2}$ (gray), CO (blue), \ion{C}{1} (green), and \ion{C}{2} (red) as a function of total density ($n_\mathrm{HI}+2n_\mathrm{H2}$) for pixels with $T_\mathrm{k}<100$\,K in Model A at (a) 0.5\,Myr and (b) 3.0\,Myr.
The black and white contours include 20\%, 40\%, 60\% and 80\% of H$_{2}$ data points and CO data points, respectively.
\label{fig:abundance}}
\end{figure}

Figure \ref{fig:abundance} shows density and relative abundance of H$_{2}$, CO, \ion{C}{1}, and \ion{C}{2} as a function of total density in Model A at 0.5\,Myr and 3.0\,Myr.
In Figure \ref{fig:abundance} majority of hydrogen is in \ion{H}{1}, and H$_{2}$ is about 0.01--0.1 of \ion{H}{1} in number.
The amount of \ion{H}{2} is negligibly small.
Model A at 0.5\,Myr, among the various epochs of Model A, matches best the UV absorption measurements of H$_{2}$ abundance and $N_\mathrm{HI}$ in the local interstellar medium \citep{2018ApJ...860...33F}.
At 0.5\,Myr, CO is a minor constituent with abundance of $\sim 10^{-10}$ at low density around 100\,cm$^{-3}$, while it grows with density to $10^{-6}$ at density above $10^{3}$\,cm$^{-3}$.
At the highest density around $(2\mbox{--}5)\times 10^{3}$\,cm$^{-3}$, CO abundance of $\sim 10^{-4}$, close to the total C abundance, is reached in a small number of pixels.
\ion{C}{1} shows an abundance increase from $10^{-7}$ to $10^{-5}$ with density in the same density range, and this variation is smaller than in CO. \ion{C}{2} shows fairly uniform abundance of $\sim 10^{-4}$ and does not depend on density.
Only at the highest end of $(\mbox{2--5})\times 10^{3}$\,cm$^{-3}$ \ion{C}{2} abundance shows slightly larger dispersion, which is due to net conversion of \ion{C}{2} into CO.
Table \ref{tab:abundance} presents a summary of the relative abundances of the species along with the CNM volume filling factor of the models.

\begin{figure}
\includegraphics{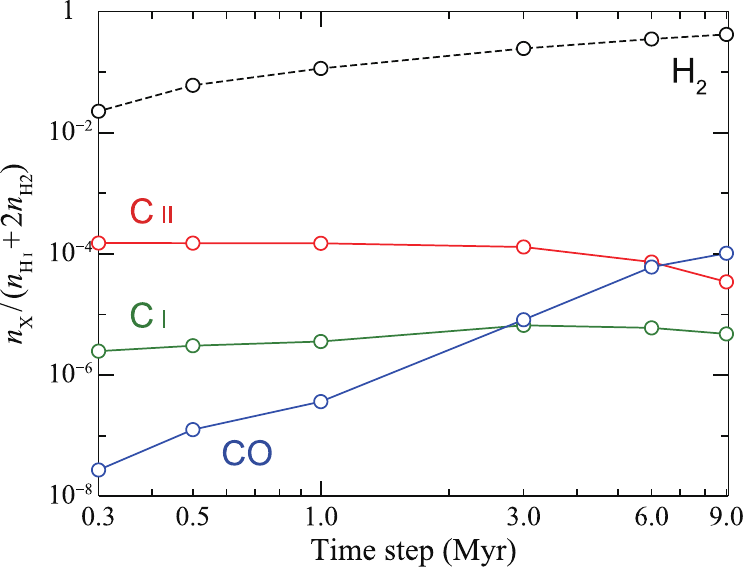}
\caption{
Time evolution of abundance of the four species of Model A measured at $10^{3}$\,cm$^{-3}$ at epochs from 0.3\,Myr to 9.0\,Myr.
\label{fig:timeevolution}}
\end{figure}

Figure \ref{fig:timeevolution} shows time evolution of abundance of the four species of Model A measured at $10^{3}$\,cm$^{-3}$ at epochs from 0.3\,Myr to 9.0\,Myr. 
Abundance of H$_{2}$, CO and \ion{C}{1} increases in time via chemical reactions including H$_{2}$ formation and due to the increase in shielding by the increase in $N_\mathrm{H}$ and $A_\mathrm{V}$.
Almost full conversion of \ion{H}{1} into H$_{2}$ is seen at 9\,Myr, while \ion{H}{1} is always dominant at epochs younger than 6\,Myr.
This dominance of \ion{H}{1} is not observed in the previous models, which for instance adopted the initial condition that \ion{H}{1} has high uniform density$=100$\,cm$^{-3}$ and static velocity field \citep{2007ApJS..169..239G}.
In Model A CO abundance shows a steep gradient in time from less than $10^{-6}$ to $10^{-4}$ over the time range.
The Model A at 9\,Myr represents a mature molecular cloud which has H$_{2}$ and CO relative abundance of 1.0 and $10^{-4}$, respectively.

\begin{figure*}
\includegraphics[scale=0.85]{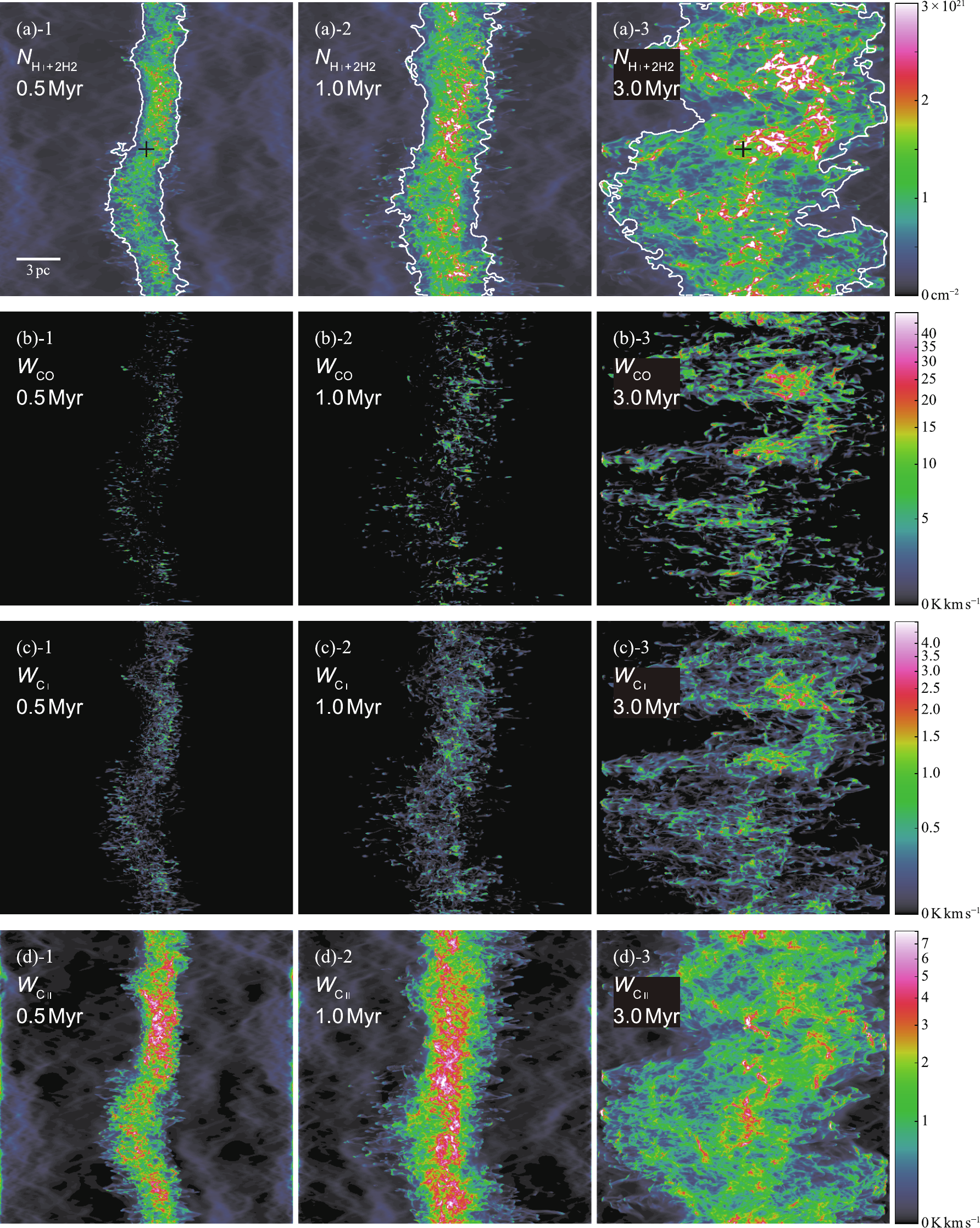}
\caption{
Spatial distribution of Model A at 0.5, 1.0 and 3.0\,Myr (from left to right);
(a) model total column density ($N_\mathrm{HI}+2N_\mathrm{H2}$), (b)--(d) velocity-integrated intensity of the synthetic observed CO, \ion{C}{1} and \ion{C}{2}.
The images are $\mbox{20\,pc}\times \mbox{20\,pc}$ in size and have a resolution of 0.04\,pc per pixel.
The $x$ and $z$ axes in the numerical domain correspond to the horizontal and the vertical axes in each panel.
\label{fig:maps}}
\end{figure*}

Figure \ref{fig:maps} shows spatial distribution of the four quantities in Model A at three epochs, 0.5, 1.0 and 3.0\,Myr.
The quantities are total hydrogen column density $N_\mathrm{HI}+2N_\mathrm{H2}$, the total velocity integrated intensity of the CO $J=\mbox{1--0}$ emission at 115.2712018\,GHz $W_\mathrm{CO}$, that of \ion{C}{1} $^{3}\textrm{P}_{1}$--$^{3}\textrm{P}_{0}$ emission at 492.160651\,GHz $W_\mathrm{CI}$, and that of \ion{C}{2} $^{2}\textrm{P}_{3/2}$--$^{2}\textrm{P}_{1/2}$ emission at 157.7409\,{\micron} $W_\mathrm{CII}$.
The distributions show marked difference between 0.5 and 1.0\,Myr vs.\ 3.0 and 6.0\,Myr.
In the early epochs the distributions are largely dominated by small clumps, while in the latter epochs the emission becomes more continuous and organized into larger aggregations.
We present an interpretation of this trend as signatures of molecular cloud formation in Section \ref{sec:discussion}. 
We note that the distribution is highly transient in a typical timescale of $10^{4}$\,yr, and any hysteresis in the Model is rapidly lost within that time scale.

\begin{figure*}
\includegraphics{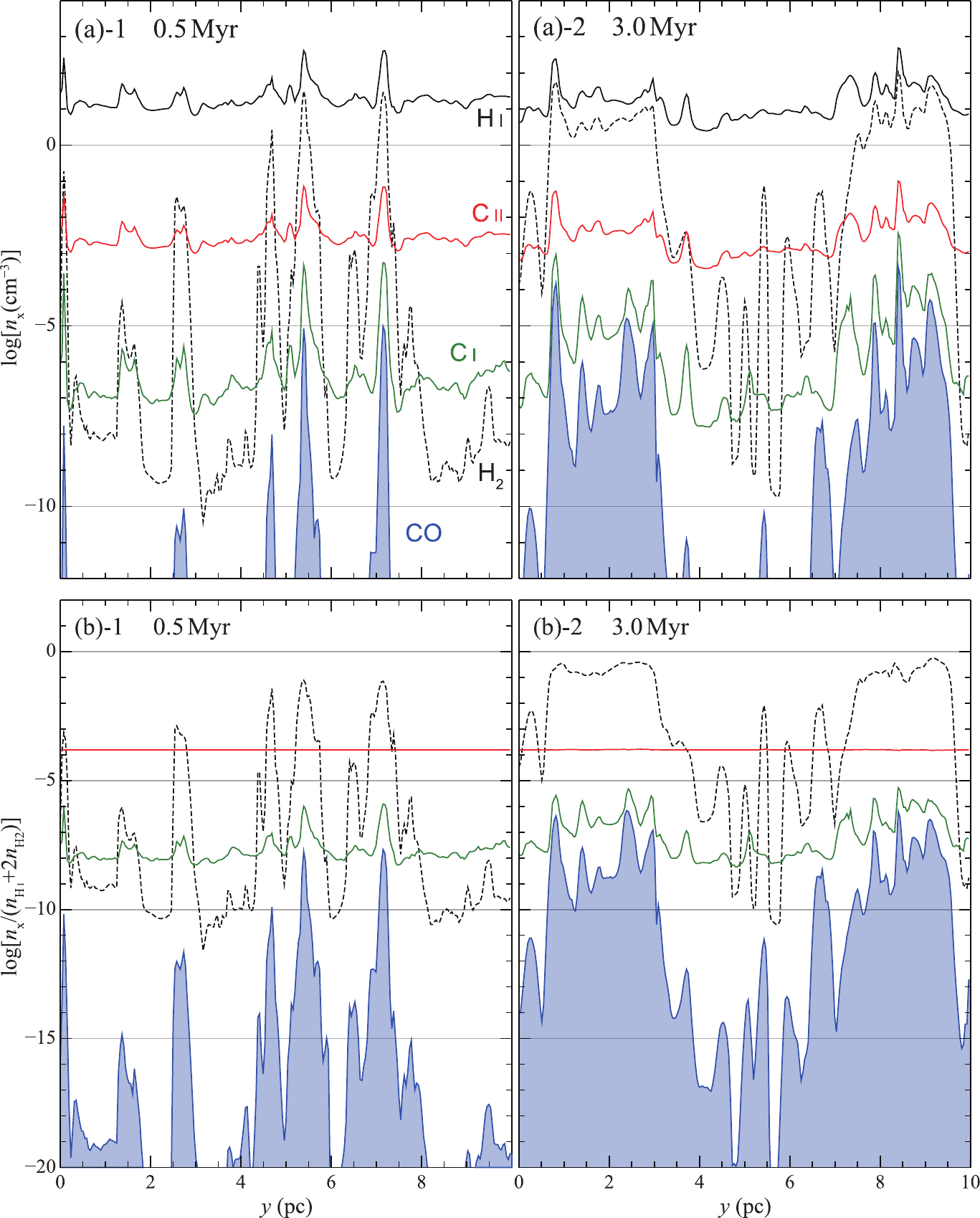}
\caption{
(a) Density $n_\mathrm{X}$ (X=\ion{H}{1}, H$_{2}$, \ion{C}{1}, \ion{C}{2} and CO) profiles in Model A at 0.5\,Myr ((a)-1) along the line of sight toward the cross mark in Figure \ref{fig:maps}(a)-1, and at 3.0\,Myr ((a)-2) toward the mark in Figure \ref{fig:maps}(a)-3.
The horizontal axes are the distance from the far side of the model ISM, $y$.
The black solid and dashed lines show \ion{H}{1} and H$_{2}$.
The red, green and blue lines show \ion{C}{2}, \ion{C}{1} and CO, respectively.
(b) Fraction $n_\mathrm{X}/(n_\mathrm{HI}+2n_\mathrm{H2})$ profiles.
\label{fig:densityprofile}}
\end{figure*}

Two sightlines are chosen rather arbitrarily in Figure \ref{fig:maps} and the strip maps of density and abundance are shown in the sightlines in Figure \ref{fig:densityprofile} at two epochs, 0.5\,Myr and 3.0\,Myr.
This allows us to see the detailed variation in space.
Figure \ref{fig:densityprofile}(a)-1 shows an early phase of evolution at 0.5\,Myr and shows dense clumps as sharp spikes with highest H$_{2}$ density above $10^{2}$\,cm$^{-3}$ at four positions, $y=2.7$\,pc, 4.6\,pc, 5.5\,pc and 7.1\,pc, where the one close to the edge, $y=0.1$\,pc, is excluded.
The clumps have peak density of 100--500\,cm$^{-3}$.
Figure \ref{fig:densityprofile}(b)-1 shows abundance of four species, H$_{2}$, \ion{C}{2}, \ion{C}{1} and CO.
H$_{2}$ abundance is peaked at $10^{-1}$ toward the two densest clumps, the 3rd and 4th, and the 2nd and 1st clumps have H$_{2}$ abundance of $\sim 7\times 10^{-2}$ and $\sim 10^{-3}$, respectively.
In Figure \ref{fig:densityprofile}(b)-1 \ion{C}{2} abundance is quite uniform at $10^{-4}$, and \ion{C}{1} abundance shows moderate variation from $10^{-8}$ to $10^{-6}$.
Compared to these, the CO abundance changes significantly from less than $10^{-20}$ to $10^{-7}$, showing maximum values toward the 3rd and 4th clumps.
Figure \ref{fig:densityprofile}(a)-2 shows a later phase at 3\,Myr in the same sightline.
We find two outstanding regions with enhanced H$_{2}$ density at $y=0.8$--3.0\,pc and 7.5--9.5\,pc, which have H$_{2}$ abundance greater than 0.1.
The general behavior of the four species is similar to the epoch of 0.5\,Myr.
The CO abundance is significantly increased to $10^{-10}$--$10^{-6}$ at 3.0\,Myr.

\subsection{Synthetic observations of CO, \ion{C}{1} and \ion{C}{2}}
\begin{figure*}
\includegraphics{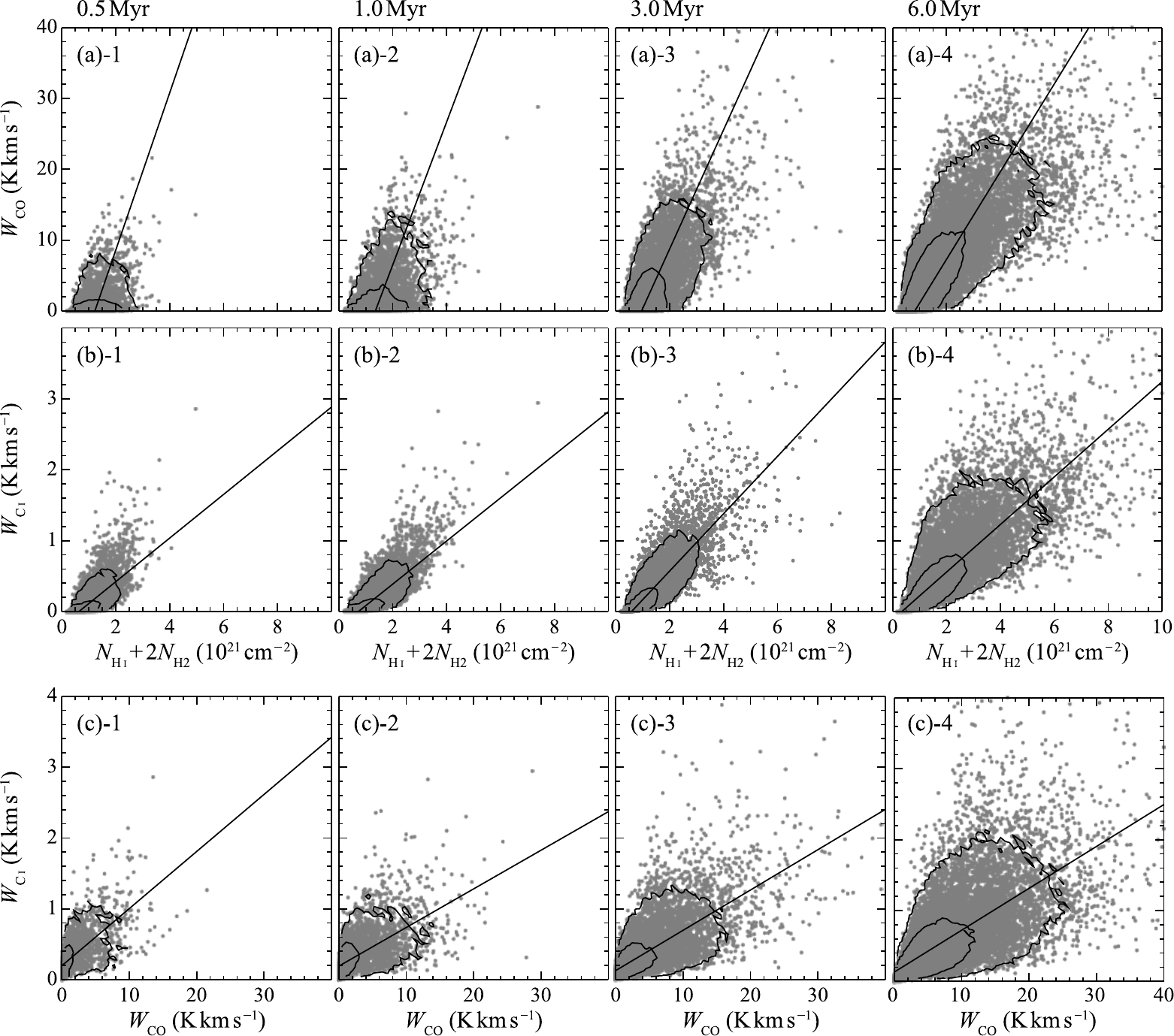}
\caption{
(a) Correlation plots of $W_\mathrm{CO}$ vs.\ $(N_\mathrm{HI}+2N_\mathrm{H2})$ in Modle I at 0.5, 1.0, 3.0 and 6.0\,Myr (from left to right).
(b) Those of $W_\mathrm{CI}$ vs.\ $(N_\mathrm{HI}+2N_\mathrm{H2})$ and (c) $W_\mathrm{CO}$ vs.\ $W_\mathrm{CI}$.
The contours include 70\% and 95\% of data points in each plot and the straight lines show linear regression line by orthogonal fitting (the slopes and dispersions are summarized in Table \ref{tab:fittingresults}).
\label{fig:corrplot}}
\end{figure*}

Figures \ref{fig:corrplot}(a), (b) and (c) show scatter plots between $W_\mathrm{CO}$ vs.\ $N_\mathrm{HI}+2N_\mathrm{H2}$, $W_\mathrm{CI}$ vs.\ $N_\mathrm{HI}+2N_\mathrm{H2}$ and $W_\mathrm{CO}$ vs.\ $W_\mathrm{CI}$.
The distributions of $W_\mathrm{CO}$ and $W_\mathrm{CI}$ in Figure \ref{fig:corrplot}(c) are similar with each other.
Each plot was fitted by an orthogonal reduced-major-axis fitting and the slopes and dispersion are summarized in Table \ref{tab:fittingresults}. 
Each plot shows positive correlations with correlation coefficients of $\sim 0.7$ except for the early phase of $W_\mathrm{CO}$ vs.\ $N_\mathrm{HI}+2N_\mathrm{H2}$.
$W_\mathrm{CO}$ is poorly correlated with $N_\mathrm{HI}+2N_\mathrm{H2}$ at 0.5 and 1.0\,Myr, reflecting very low CO abundance (see Table \ref{tab:abundance})

In Figure \ref{fig:corrplot} the intensities, $W_\mathrm{CO}$ and $W_\mathrm{CI}$, are less than 20\,K\,km\,s$^{-1}$ and 1.3\,K\,km\,s$^{-1}$, respectively, and the corresponding column density is less than $4\times 10^{21}$\,cm$^{-2}$ at epochs earlier than 1.0\,Myr.
We note that in the epoch 0.5\,Myr, where the CO abundance is $10^{-6}$, $W_\mathrm{CO}$ becomes $\sim 10$\,K\,km\,s$^{-1}$ and is observable by the current instruments.
As seen in Table \ref{tab:abundance} the CO clouds in this phase is still dominated by \ion{H}{1}, and H$_{2}$ abundance is about 10\%. 

\section{Discussion}\label{sec:discussion}

\subsection{Numerical sumulations of the \ion{H}{1} distribution and comparison with the \ion{H}{1} and dust observations}
\citet{2012ApJ...759...35I} presented H$_{2}$ formation in converging \ion{H}{1} flows which models the local interstellar \ion{H}{1} gas, and calculated the three dimensional distribution of H$_{2}$, CO, \ion{C}{1} and \ion{C}{2} as well as the gas density and pressure distributions at 0.02\,pc resolution over a timescale of 10\,Myr. 
Similar numerical simulations to \citet{2012ApJ...759...35I} were made by the other authors including \citet{2007ApJS..169..239G, 2008A&A...486L..43H, 2014A&A...571A..46V, 2016A&A...587A..76V}. \citet{2007ApJS..169..239G} assumes the CNM only in the initial state of the simulations. These authors calculated both for the static and turbulent initial velocity fields, and showed rapid molecular formation in the turbulent field. This work is different from \citet{2012ApJ...759...35I} in that the UV shielding is effective from the beginning of the simulations due to the high density CNM. In the simulations by \citet{2008A&A...486L..43H}, the chemical network is not incorporated, which is the major difference from \citet{2012ApJ...759...35I}. \citet{2014A&A...571A..46V, 2016A&A...587A..76V} made follow up simulations with chemical reactions. \cite{2016A&A...587A..76V} argued that H$_2$ formation is rapid as compared with the static case, and found that \ion{H}{1} is dominant as compared with H$_2$ in the early stage of evolution with $t<10$ Myr. This trend is consistent with \citet{2012ApJ...759...35I}.
\citet{2013ApJ...776....1K} made large-scale \ion{H}{1} simulations in a galactic kpc-scale with lower spatial resolution of 2 pc, insufficient to resolve thermal instabilities.

\citet{2012ApJ...759...35I} focused on the analysis of the basic gas physical properties such as energy and velocity, and showed the general trend of slow H$_{2}$ formation on dust surfaces and the dominant \ion{H}{1} fraction over 10\,Myr after the initiation of the \ion{H}{1} flow collision.
A confrontation of the predicted gas distributions with the observations was, however, beyond the scope of \citet{2012ApJ...759...35I}, and interpretation of the observations of H$_{2}$, CO and \ion{C}{1} in a framework of \citet{2012ApJ...759...35I} remained as a task for better understanding molecular cloud formation observationally. 

As a step of applying the \citet{2012ApJ...759...35I} results to the observations, \citet{2018ApJ...860...33F} made a detailed visualization of the spatial distribution of \ion{H}{1} and H$_{2}$ and compared the synthetic observational results with the observations of \ion{H}{1}, the emission-absorption measurements toward radio continuum sources, and UV absorption measurements of H$_{2}$.
An important outcome of \citet{2018ApJ...860...33F} was the highly clumped distribution of the CNM, which occupies only 3--4\% of the total volume.
Such an extremely small filling factor of the CNM was not explicitly mentioned often in the literature, while qualitatively a small filling factor of the CNM was suggested previously \citep[e.g.,][]{2003ApJ...586.1067H}.
A usual understanding of the CNM and WNM is that the two have comparable mass and that the CNM is by a factor of 30 denser than the WNM.
The $\sim 30$ times difference in average density between the CNM and WNM leads immediately to robust knowledge that the CNM volume filling factor is 3--4\% of the WNM \citep[e.g., see Table 1 of][]{2011piim.book.....D}, and this is consistent with the synthetic observations of \citet{2018ApJ...860...33F}.
As a general remark we need to be cautioned to use a ``typical'' value in calculating \ion{H}{1} properties because the \ion{H}{1} density and temperature have broad ranges over a few orders of magnitude with no clear peaks (Figure \ref{fig:histo_n+T}).
Some typical value of \ion{H}{1} thus sometimes can oversimplify the \ion{H}{1} properties, and careful testing by considering distribution functions of the relevant parameters are required.

It could be questioned if the results of \citet{2012ApJ...759...35I} may be a special situation caused by the continuous \ion{H}{1} inflow which may produce too high pressure and too high CNM clumpiness \citep[e.g.,][]{2018ApJ...862..131M}.
In order to test a possible model bias, we used several different models in the present work (Section \ref{sec:models}).
The new Models with different setups show a small volume filling factor of the CNM (Section \ref{sec:models} and Table \ref{tab:abundance}), lending further support for the CNM small filling factor as a general property of the interstellar \ion{H}{1}.
It is also to be noted that the gas distribution in the present simulations are all transient and is not in equilibrium.
The clumps are not in pressure balance but changes in time by a typical timescale of 0.1\,pc/ 3\,km\,s$^{-1}$ $\sim$ a few times $10^{4}$\,yr, and Figure \ref{fig:maps} presents snapshots at each epoch.
In this connection, the pressure of the ISM measured in the \ion{C}{1} absorption \citep{2011ApJ...734...65J} offers another possible test of the models.
These authors derived $\log(p/k_\mathrm{b})=3.58\pm 0.175$ dex (rms) as the average pressure of the ISM based on the UV absorption of the ground state \ion{C}{1}, where the rms error is a lower limit. Into more detail, their Figures 7 and 8 show a distribution function of $\log(p/k_\mathrm{b})$; all the data of $\log(p/k_\mathrm{b})$ have a broad range from 2.5 to 4.5 and the major portion is distributed from 3.0 to 4.2 for $dN(\mathrm{H})/d(\log(p/k_\mathrm{b}))$ above $10^{22}$ cm$^{-2}$.
In the present model, \ion{C}{1} samples mainly the low-density end, $\sim 10^{2}$\,cm$^{-3}$, of the CNM and its pressure weighted by \ion{C}{1} mass is from 3.7 to 4.2, which is somewhat higher but is not inconsistent with the result of \citet{2011ApJ...734...65J}.

\citet{2018ApJ...860...33F} argued that the emission-absorption measurements usually underestimate the \ion{H}{1} column density because the measurements bias toward the WNM which has a covering factor of 80\% in the sky.
Most of the emission-absorption measurements sample the optically thin WNM.
Since the WNM has significantly small \ion{H}{1} optical depth of $\sim 10^{-2}$ due to high spin temperature above 1000\,K, the emission-absorption measurements biased toward the WNM underestimate the \ion{H}{1} column density $N_\mathrm{HI}$.
On the other hand, the \textit{Planck}/\textit{IRAS}-based measurements \citep{2014ApJ...796...59F,2015ApJ...798....6F} derived larger $N_\mathrm{HI}$ than the emission-absorption measurements.
The dust grains are included both in the CNM and WNM, and the dust emission samples the whole ISM volume.
This difference in observational bias in the emission-absorption measurements explains the difference from the dust-based measurements.
\citet{2018ApJ...860...33F} in addition warned the extremely small area covering factor subtended by the radio continuum point-like sources used in the emission-absorption measurements which has density of one source every 10 square degrees in the sky, and argued that the emission-absorption measurements are not suited to derive bulk \ion{H}{1} properties in the local ISM.
A remaining issue to be better understood in the context is the dust property which was used as a proxy of hydrogen in the \textit{Planck}/\textit{IRAS} based method.
Possible changes in dust emission properties need to be further carefully worked out. In this connection, \citet{2018ApJ...860...33F} showed that the small dispersion of the optically thin regime, in the scatter plot between $W_\mathrm{HI}$ and $\tau_{353}$, the dust optical depth at 353\,GHz, is consistent with the uniform dust property, whereas a range of the dust optical depth is limited.
It is noteworthy that, most recently, the dust optical depth as a proxy of the ISM is successfully applied to measure metallicity in the ISM in the Galactic halo and the Local Group galaxies \citep{2017PASJ...69L...5F,2017arXiv171109529F}.

\subsection{The seed CO clouds and their observations}
In the classical framework of molecular cloud formation, the density inhomogeneity and turbulence were not often considered in theoretical works \citep{2014prpl.conf....3D}.
The present results, incorporating fully the turbulent inhomogeneities, show that the CNM distribution places the initial setup for H$_{2}$ formation, and H$_{2}$ formation proceed first in the heart of the CNM clumps in a time scale shorter than 1\,Myr.
We note that the H$_{2}$ abundance in the earliest phase $\lesssim 1$\,Myr is as low as $10^{-1}$ at maximum in the CNM clumps and almost full conversion of \ion{H}{1} into H$_{2}$ is realized at density 100\,cm$^{-3}$ in 10\,Myr (Figure \ref{fig:timeevolution}).
In a timescale less than 3\,Myr CO clouds are dominated by \ion{H}{1}, and \ion{H}{1} occupies more than 70\% of hydrogen (see Figures \ref{fig:densityprofile}(a)-1 and \ref{fig:densityprofile}(b)-1 and Table \ref{tab:abundance}).
Such a picture of \ion{H}{1} dominance in young CO clouds derived from the simulations by \citet{2012ApJ...759...35I} is not usual in the literature.
\citet{2018ApJ...860...33F} showed that the CNM is distributed in small clumps whose size is in a sub-pc scale. The early spatial distribution of the CO clouds shows naturally highly clumpy distribution similar to the CNM \citep{2012ApJ...754...95T,2018ApJ...860...33F}. 

The present results predict the appearance of young CO clouds as clumpy \ion{H}{1}-dominated clouds of sub-pc size with low CO abundance of $10^{-8}$ to $10^{-6}$ even at their peak.
Possible evidence for the seed CO cloud formation was reported in the Pegasus loop where an early B2 star HD 866 acted to compress the ambient \ion{H}{1} gas into a shell of 30\,pc radius at 250\,pc distance (\citealt{2006ApJ...642..307Y}; distance 100\,pc assumed by these authors is revised to 250\,pc by Saeki et al.\ in prep.\ based on the new star count data).
Small CO clouds of sub-pc to pc scales are distributed without massive CO cloud as revealed by the large scale CO $J=1$--0 observations with NANTEN \citep{2006ApJ...642..307Y}.
\citet{2006ApJ...642..307Y} paid attention to that the clouds show weak CO peak intensities of a few K which is unusual among the typical local dark clouds, and discussed that they may represent young CO clouds forming in shock compression by the B2 star within $\sim 1$\,Myr.
These authors compared the data with the cloud formation theory of \citet{2002ApJ...564L..97K}, which is the earlier work of \citet{2012ApJ...759...35I}.
In the Pegasus region the cloud mass ranges from 0.1\,$M_\sun$ to 50\,$M_\sun$ with a median of 5\,$M_\sun$, and most of them have radius less than 0.6\,pc.
These properties are consistent with the seed CO clouds in the present work.

Another region of similar small clouds in Ophiuchus was studied by \citet{2012ApJ...754...95T}.
These authors focused on a CO cloud named Cloud S (also known as \object{LDN 204}) in the Ophiuchus North region around an O9.5 star \object{$\zeta$ Oph}.
The cloud was mapped in the CO $J=1$--0 emission by the Nagoya 4\,m mm-telescope \citep{1991ApJS...77..647N} and NANTEN \citep{2000PASJ...52.1147T}, and follow up higher resolution studies were made with the Nobeyama Radio Observaotry 45\,m+BEARS receiver and ASTE 10\,m telescopes in CO 1--0 and 3--2 emissions \citep{2012ApJ...754...95T}.
These results showed small $^{12}$CO cloudlets having $\mbox{linewidth}=0.6$\,km\,s$^{-1}$, $\mbox{radius}=3000$--6300\,AU, $\mbox{mass}=0.005$--0.05\,$M_\sun$, $\mbox{density}=(\mbox{1--11})\times 10^{3}$\,cm$^{-6}$ and high $T_\mathrm{k}=25$--300\,K irradiated by the O star.
Most recently, new ALMA observation further resolved $^{12}$CO clouds having $\mbox{linewidth}=0.3$\,km\,s$^{-1}$, $\mbox{radius} \sim1000$\,AU, and $\mbox{mass}\sim 10^{-4}$\,$M_\sun$, in Cloud S (Tachihara et al.\ in prep.).
These results confirm very small scale distributions in the CO clouds, lending further support for the small structures predicted by the present study based on \citet{2012ApJ...759...35I}, while the ALMA resolution is higher than that of the current numerical simulations. 

The present study also offers an insight into the later phase of CO cloud formation in particular at epochs after 3\,Myr.
The synthetic distributions suggest that small CO clouds likely merge into a large CO cloud by coagulation in a timescale longer than 3\,Myr (Figure \ref{fig:densityprofile}).
A scenario of aggregation of small molecular clouds as a formation mechanism of large molecular clouds was presented theoretically (e.g., \citealt{2014prpl.conf....3D}; see also \citealt{2001MNRAS.327..663P}), whereas we are not aware of detailed numerical studies like the present one on the aggregation.

\subsection{Similar pc-scale distributions of CO and \ion{C}{1}}
Some early \ion{C}{1} data were taken by \citet{1999ApJ...524L.129M,1999ApJ...527L..59I,2003ApJ...589..378K} in local dark clouds and some more distant GMCs by \citet{2004A&A...424..887K}, \citet{2014ApJ...782...72B} and \citet{2013ApJ...774L..20S}.
A common result of these studies is that two distributions of \ion{C}{1} and CO are similar with each other in spite of that \ion{C}{1} is formed in the less $A_{V}$ layers than CO as shown by the PDR layered model by \citet{1985ApJ...291..722T}.
According the PDR model, the \ion{C}{1} layer with \ion{H}{1} is at less than $A_{V}=2$\,mag and H$_{2}$ becomes dominant at $A_{V}$ more than 2 mag where CO is still not abundant due to the less self-shielding of CO than H$_{2}$.
The CO-dark H$_{2}$ layer is distributed at $A_{V}$ between 2 mag and 3 mag \citep{1985ApJ...291..722T}.
H$_{2}$ with a CO layer follows this at more than $A_{V}=3$\,mag.
$A_{V}=1$\,mag corresponds to $N_\mathrm{HI}=1.9\times 10^{21}$\,cm$^{-2}$ mag and in the layered model, the mass of the CO-dark H$_{2}$ may be significant.
In the current picture of highly clumped \ion{C}{1} and CO distribution, the CO-dark H$_{2}$ is in the surface of each clump and $A_{V}$ to protect H$_{2}$ is a combination of $A_{V}$ in the extended WNM and in the thin shell of the CNM clump.
As a result, the mass fraction of CO-dark H$_{2}$ is negligibly small as shown in Table \ref{tab:massbudget}.
It is not likely that the CO-dark H$_{2}$ is a major mass carrier in the local ISM, corresponding to only 1--2\% of the total ISM mass.
This is consistent with the similar spatial distribution of \ion{C}{1} and CO observed in various clouds \citep{2014ApJ...782...72B}.
The work by \citet{2015MNRAS.448.1607G} also explained the similar \ion{C}{1} and CO distribution by a turbulent model which allows UV radiation to penetrate deeply into a cloud.
The clumpy distribution of CO and \ion{C}{1} provides a basis for the similarity between the CO and \ion{C}{1} distributions averaged at a pc-scale.
The present model shown in Figure \ref{fig:densityprofile} has $N_\mathrm{HI}+2N_\mathrm{H2}=(1\mbox{--}1.5)\times 10^{21}$\,cm$^{-2}$, $N_\mathrm{HI, WNM}=(0.3\mbox{--}0.5)\times 10^{21}$\,cm$^{-2}$, and $N_\mathrm{HI, CNM}=(0.2\mbox{--}1.1)\times 10^{21}$\,cm$^{-2}$ \citep[see also Table 5 of][]{2018ApJ...860...33F}.
This sightline does not include the peak of a CNM clump, where $A_{V}$, a sum of the WNM and CNM, can become as large as more than 2 mag which corresponds to $A_{V}$ for the \ion{H}{1}/H$_{2}$ transition.
Because $A_{V}$ of the WNM is somewhat smaller than that of the CNM, UV radiation penetrates more deeply in the clumpy case than the layered case, allowing to produce \ion{C}{1} throughout the ISM.

\ion{C}{1} and CO are not extremely accurate tracers of the ISM as indicated by the relatively low correlation coefficients in the scatter plots $W_\mathrm{CO}$ vs.\ $N_\mathrm{HI}+2N_\mathrm{H2}$ or $W_\mathrm{CI}$ vs.\ $N_\mathrm{HI}+2N_\mathrm{H2}$ in Figure \ref{fig:corrplot}.
For a given $W_\mathrm{CO}$ or $W_\mathrm{CI}$, $N_\mathrm{HI}+2N_\mathrm{H2}$ is constrained typically within a factor of 2 as shown in Section \ref{sec:results}.
This large range is not inconsistent with the usual error range of a factor of 2 in $X_\mathrm{CO}$, a conversion factor of $W_\mathrm{CO}$ into $N_\mathrm{H2}$ \citep{2017ApJ...838..132O}.
In the present Model setup, this is ascribed to the large variation of their abundance which changes with the local sub-pc scale conditions including UV radiation in each CNM clump.
It is also noteworthy that \ion{H}{1} is more abundant than H$_{2}$ at an age shorter than 10\,Myr.
Only after 10\,Myr full conversion of \ion{H}{1} into H$_{2}$ is realized where the \ion{H}{1} abundance $\sim 10^{-2}$ \citep[e.g.,][]{1978AJ.....83.1607S} is primarily determined by the cosmic ray dissociation of H$_{2}$.
The \ion{H}{1} abundance in a CO cloud in its early evolutionary stages is dominated by non-equilibrium chemical reactions.

\subsection{CO-dark H$_{2}$}
Table \ref{tab:massbudget} summarizes the mass budget of the ISM, \ion{H}{1} and H$_{2}$, as probed by the usual tracers including CO and \ion{C}{1} in the present study.
In particular, the CO-dark H$_{2}$ has attracted attention as a possible important mass carrier in the ISM.
It is argued that there is a significant amount of H$_{2}$ gas which is not detectable in CO emission.
CO is dissociated by UV radiation at 11.2\,eV, while H$_{2}$ is dissociated by UV radiation at 4.5\,eV.
H$_{2}$ is however more heavily protected than CO by self-shielding by its high abundance and a layer is formed where only H$_{2}$ is abundant and CO is much less, and undetectable.
A theoretical model of the envelope of a GMC was presented by \citet{2010ApJ...716.1191W} and showed that CO-dark H$_{2}$ has a significant fraction of gas mass.
This model is however not aimed at simulating the local ISM which has more turbulent and transient properties in the solar neighorhood.
In particular, the timescale of the model by \citet{2010ApJ...716.1191W} is $\sim 10$\,Myr and chemical equilibrium is assumed.
This is not the case in the local ISM which is often disturbed by the SNR shocks in a 1\,Myr scale as argued by \citet{2002ApJ...564L..97K,2004ApJ...602L..25K}.
It is possible that CO-dark H$_{2}$ can be important in the GMCs, whereas it is not applicable to the transient local ISM which is the subject of the present study.
The results in Table \ref{tab:massbudget} show that the H$_{2}$ mass fraction is small, 10--30\% at epochs 1--3\,Myr and the fraction of the CO-dark H$_{2}$ is 10\% of the total H$_{2}$ and only 1--2\% of the total ISM.
The CO-dark H$_{2}$ is therefore not likely as a dark mass candidate in the local ISM.

\section{Conclusions}\label{sec:conclusions}

We carried out synthetic observations of the interstellar carbon in the forms of CO, \ion{C}{1} and \ion{C}{2}.
The theoretical model based on magneto-hydrodynamical numerical simulations of \ion{H}{1} gas converging at a velocity of 20\,km\,s$^{-1}$ by \citet{2012ApJ...759...35I} is employed.
The gas is turbulent and inhomogeneous with magnetic field.
The model parameters at an epoch of 0.5\,Myr were chosen to be consistent with the observations of H$_{2}$ in UV and 21\,cm \ion{H}{1} emission-absorption measurements of the local interstellar medium around the sun.
We focus on the evolution of the rare species, CO, \ion{C}{1} and \ion{C}{2}, other than the main constituents H$_{2}$ and \ion{H}{1} from 0.3\,Myr to 9\,Myr since the converging flow initiated interaction.
We find two phases of cloud formation. In the early phase within 1\,Myr, the CNM clumps of sub-pc scales are the formation sites of H$_{2}$, \ion{C}{1} and CO, and in the late phase from 3\,Myr to 9\,Myr the CNM clumps merge together to form pc-scale molecular clouds.
In the early phase the clumps are dominated by cool \ion{H}{1} and the typical H$_{2}$ fraction is $\sim 10$\% of \ion{H}{1}, and the abundances of \ion{C}{1} and CO are $\sim 10^{-5}$ of H$_{2}$.
This predicts that the CO clumps in the early phase, the seed CO clouds, are detectable as sub-pc scale weak CO clouds.
In the late phase the merged molecular clouds are dominated by H$_{2}$ where $n_\mathrm{H2}/(n_\mathrm{HI}+2n_\mathrm{H2})\sim 0.1$--1.0 and the CO abundance reaches $\sim 10^{-4}$.
These results are consistent with the large-scale observations of \ion{C}{1} and CO, both of which show fairly similar spatial distributions due to the sub-pc clumpy distribution of \ion{C}{1} and CO emitting regions.
This makes a significant contrast with the classic picture of the PDR which predicted distinct pc-scale layers of \ion{C}{1} and CO in the sequence of $A_{V}$ \citep{1985ApJ...291..722T}.
We identified observational signatures of the molecular clouds in their infancy by low CO and high \ion{H}{1} abundance where H$_{2}$ is yet minor species.
We suggest that the faint CO clouds in the Ophiucus North \citep{2012ApJ...754...95T} and the Pegasus loop \citep{2006ApJ...642..307Y} are good candidates for such an early phase.
We also note that the fraction of CO-dark H$_{2}$ is small, less than a few percent, and not important in the total mass budget of the local ISM which is dominated by the \ion{H}{1}.
This disagrees with the conventional idea that CO-dark H$_{2}$ is one of the major mass carriers in the ISM.

\acknowledgments
This work was supported by JSPS KAKENHI Grant Number JP15H05694.

\bibliography{references}

\clearpage

\clearpage

\begin{deluxetable*}{lRRRRRRRRR}
\tablecaption{Chemical abundance and the CNM volume filling factor\label{tab:abundance}}
\tablehead{
 & \multicolumn6c{Model A} & \colhead{Model B} & \colhead{Model C} & \colhead{Model D} \\
\cline{2-7}
 & \colhead{0.3\,Myr} & \colhead{0.5\,Myr} & \colhead{1.0\,Myr} & \colhead{3.0\,Myr} & \colhead{6.0\,Myr} & \colhead{9.0\,Myr}
 & \colhead{0.5\,Myr} & \colhead{0.5\,Myr} & \colhead{0.5\,Myr}
}
\startdata
\multicolumn{10}{l}{$\log\left[n_\mathrm{X}/(n_\mathrm{HI}+2n_\mathrm{H2})\right]$\tablenotemark{a} (X=H$_{2}$, \ion{C}{2}, \ion{C}{1}, CO)} \\
\hline
H$_{2}$ & -1.6 & -1.2 & -0.9 & -0.6 & -0.5 & -0.4 & -1.3 & -1.2 & -1.4 \\
\ion{C}{2} & -3.8 & -3.8 & -3.8 & -3.9 & -4.1 & -4.5 & -3.9 & -3.9 & -3.9 \\
\ion{C}{1} & -5.6 & -5.5 & -5.4 & -5.2 & -5.2 & -5.3 & -4.6 & -4.8 & -4.8 \\
CO & -7.6 & -6.9 & -6.4 & -5.0 & -4.2 & -4.0 & -5.1 & -5.4 & -5.9 \\
\hline
\multicolumn{10}{l}{CNM volume filling factor} \\
\hline
 & 3.4\% & 3.5\% & 5.0\% & 12.6\% & 29.0\% & 37.2\% & 4.4\% & 6.5\% & 2.4\%
\enddata
\tablenotetext{a}{Decimal logarithm of median relative abundance, measured in a $(n_\mathrm{HI}+2n_\mathrm{H2})$ range from $10^{2.9}$ to $10^{3.1}$\,cm$^{-3}$.}
\end{deluxetable*}

\begin{deluxetable*}{rRRR}
\tablecaption{Fitting results in the scatter plots of $W_\mathrm{CO}$, $W_\mathrm{CI}$ and $(n_\mathrm{HI}+2n_\mathrm{H2})$\label{tab:fittingresults}}
\tablehead{
\colhead{Time step} & \colhead{$A$} & \colhead{$B$} & \colhead{C.C.} \\
\colhead{(Myr)} 
}
\colnumbers
\startdata
\multicolumn{4}{l}{$W_\mathrm{CO} (\mbox{K\,km\,s$^{-1}$})=A(N_\mathrm{HI}+2N_\mathrm{H2})(10^{21} \mbox{cm$^{-2}$})+B$} \\
\hline
0.5 & 11.1\phn\pm 0.5\phn & -13.6\phn\pm 0.6\phn & 0.36 \\ 
1.0 & 10.1\phn\pm 0.2\phn & -13.7\phn\pm 0.3\phn & 0.48 \\
3.0 & 8.40\pm 0.06 & -8.05\pm 0.07 & 0.71 \\ 
6.0 & 6.19\pm 0.03 & -5.07\pm 0.04 & 0.76 \\
\hline
\multicolumn{4}{l}{$W_\mathrm{CI} (\mbox{K\,km\,s$^{-1}$})=A(N_\mathrm{HI}+2N_\mathrm{H2})(10^{21} \mbox{cm$^{-2}$})+B$} \\
\hline
0.5 & 0.308\pm 0.005 & -0.195\pm 0.005 & 0.67 \\
1.0 & 0.304\pm 0.003 & -0.222\pm 0.003 & 0.76 \\
3.0 & 0.404\pm 0.002 & -0.238\pm 0.002 & 0.82 \\
6.0 & 0.336\pm 0.002 & -0.115\pm 0.003 & 0.79 \\
\hline
\multicolumn{4}{l}{$W_\mathrm{CI} (\mbox{K\,km\,s$^{-1}$})=AW_\mathrm{CO} (\mbox{K\,km\,s$^{-1}$})+B$} \\
\hline
0.5 & 0.081\phn\pm 0.002\phn & 0.198\pm 0.003 & 0.69 \\
1.0 & 0.0545\pm 0.0009 & 0.191\pm 0.002 & 0.67 \\ 
3.0 & 0.0568\pm 0.0004 & 0.129\pm 0.001 & 0.73 \\
6.0 & 0.0591\pm 0.0003 & 0.133\pm 0.001 & 0.70 \\
\enddata
\tablecomments{
Columns: (2) slope and (3) intercept of orthogonal reduced-major-axis regression line,
(4) correlation coefficient.
}
\end{deluxetable*}

\begin{deluxetable*}{rrrrrrr}
\tablecaption{Mass budget\label{tab:massbudget}}
\tablecolumns{7}
\tablewidth{0pt}
\tablehead{
\colhead{Time step} & \colhead{$M_\mathrm{HI}+M_\mathrm{H2}$} & \colhead{$M_\mathrm{HI}$} & \colhead{$M_\mathrm{HI}^\mathrm{thin}$} & \colhead{$M_\mathrm{H2}$} & \colhead{$M_\mathrm{H2}^\mathrm{CO}$} & \colhead{$M_\mathrm{H2}^\mathrm{CO\mathchar"712D free}$} \\
\colhead{(Myr)} & \colhead{($M_\sun$)} & \colhead{($M_\sun$)} & \colhead{($M_\sun$)}  & \colhead{($M_\sun$)} & \colhead{($M_\sun$)} & \colhead{($M_\sun$)}
}
\colnumbers
\startdata
0.3 & 269 & 265 & 205 & 5 & 2 & 3\\
 & & [98\%] & [76\%] & [2\%] & [1\%] & [1\%] \\
0.5 & 413 & 394 & 310 & 19 & 12 & 7 \\
 & & [95\%] & [75\%] & [5\%] & [3\%] & [2\%] \\
1.0 & 820 & 746 & 564 & 75 & 64 & 11 \\
 & & [91\%] & [69\%] & [9\%] & [8\%] & [1\%] \\
3.0 & 2392 & 1732 & 1115 & 661 & 613 & 48 \\
 & & [72\%] & [47\%] & [28\%] & [26\%] & [2\%] \\ 
6.0 & 4754 & 2324 & 1191 & 2430 & 2401 & 29 \\
 & & [49\%] & [25\%] & [51\%] &  [51\%] & [1\%] \\
9.0 & 6406 & 2028 & 992 & 4377 & 4367 & 10 \\
 & & [32\%] & [15\%] & [68\%] & [68\%] & [0.1\%]
\enddata
\tablecomments{
Columns: (2) total mass of \ion{H}{1} and H$_{2}$, (3) mass of \ion{H}{1} gas, (4) mass of \ion{H}{1} given from \ion{H}{1} integrated intensity under the optically thin assumption, (5) mass of H$_{2}$, (6) mass of H$_{2}$ detected by the synthetic CO observation (see Section \ref{sec:models}) with detection limit of $W_\mathrm{CO}>0.1$\,K\,km\,s$^{-1}$, (7) mass of H$_{2}$ not detected by the synthetic CO observation.
}
\end{deluxetable*}



\end{document}